\documentclass[twocolumn,aps,pra,colorlinks,citecolor=blue, linkcolor = blue,anchorcolor = blue,urlcolor = blue,floatfix]{revtex4-2}
\usepackage{epsfig}
\usepackage[english]{babel}
\usepackage{latexsym}
\usepackage{graphics}
\usepackage{subfigure}
\usepackage{epsfig}
\usepackage{graphicx}
\usepackage{dcolumn}
\usepackage{amsmath}
\usepackage{hyperref}
\usepackage{amssymb}

\begin{document}

\preprint{APS/123-QED}

\title{Ionization phase-shifts between two laser-dressed states}

%\author{Dianxiang Ren,$^{1}$ Chao Chen,$^{2}$ Xiaokai Li,$^{1}$ Xiaoge Zhao,$^{1}$ Xitao Yu,$^{1}$ Xinning Zhao,$^{1}$   Pan Ma,$^{1}$  Chuncheng Wang,$^{1}$ Sizuo Luo,$^{1,*}$ Yanjun Chen$^{2,\dag}$ and Dajun Ding$^{1,\ddag}$}

\author{Wankai Li} 
\author{Yixuan Wang}
\author{Xing Li}
\author{Tao Yang}
\author{Dongdong Zhang}\email{dongdongzhang@jlu.edu.cn} 
\author{Dajun Ding}\email{dajund@jlu.edu.cn}

\affiliation{Institute of Atomic and Molecular Physics, Jilin University, Changchun 130012, China}
\affiliation{Jilin Provincial Key Laboratory of Applied Atomic and Molecular Spectroscopy, Jilin University, Changchun 130012,  China}
%2.College of Physics and Information Technology, Shaan'xi Normal University, Xi'an 710119, China}
\pacs{33.80.Rv, 42.65.Ky}
\date{\today}

\begin{abstract}
   Resonance-enhanced multiphoton ionization (REMPI) in potassium atom is investigated within the strong coupling regime using photoelectron momentum imaging techniques. The kinetic energy distribution of the ionized electrons reveals the eigenenergies of the dressed states, which exhibit Autler-Townes (AT) splitting. This splitting is proportional to the laser field strength. Partial wave analysis of the photoelectrons angular distributions (PAD) uncovers relative phase shifts between the dressed states, shedding light on attosecond ionization time delays. The observed phase shift between the two electron wave packets generated by the AT splitting arises from the combined effects of the Coulomb phase and the quantum defect phase.
\end{abstract}

\maketitle
\section{Introduction}
The photoelectron wave function fully characterizes the state of ionized electrons and encodes detailed information about light-matter interactions~\cite{einstein_photoelectron}. Generally, the photoelectron wave function is represented as a coherent superposition of multiple continuum states~\cite{schuricke2013coherence,PhysRevLett.60.1494}. These states are described by partial waves whose amplitudes correspond to ionization probabilities, while their relative phases contain interference information reflecting the coherence of electrons in different states. The generation, control, and measurement of photoelectron wave packets are fundamental to understanding light-matter interactions and are pivotal in optoelectronic applications. Consequently, they have been the focus of extensive research over the past decades~\cite{ohmori2009wave,PhysRevLett.99.093001}.

When a strong laser field couples two quantum states, periodic coherent population oscillations, known as Rabi oscillations, can occur~\cite{PhysRev.55.526}. A distinctive feature of such coherent interactions is the emergence of a doublet structure due to the AC-Stark effect, also known as the Autler-Townes (AT) splitting~\cite{autler1955stark}. The energy separation of the AT doublet is determined by the Rabi frequency~\cite{autler1955stark,knight1980rabi}, which reflects the strength of light-matter coupling~\cite{KNIGHT198021}. This frequency is directly influenced by the driving laser intensity, detuning, and transition dipole moment.
Rabi oscillations are crucial for quantum control and quantum information processing~\cite{RevModPhys.83.331,RevModPhys.90.025008,RevModPhys.92.015001}. For example, coherent population transfer between quantum states, governed by Rabi oscillations, is a powerful tool for exploring quantum dynamics. This principle has enabled diverse applications, including ultrafast manipulation of Rydberg states~\cite{avanaki2016harmonic,fushitani2016femtosecond,PhysRevLett.100.113003}, precision in atomic clock technology~\cite{yu2018stabilizing}, advancements in cold atom physics~\cite{PhysRevLett.100.253001,PhysRevLett.100.113003,zhang2014autler}, and control over metallic nanomaterials~\cite{vasa2013real}. The advent of coherent high-intensity XUV pulsed sources has further expanded the observation of Rabi oscillations into the extreme ultraviolet (XUV) regime~\cite{PhysRevA.84.053419,PhysRevLett.106.193008},paving the way for the exploration of novel physical phenomena.

Alkali metal atoms possess a single outermost $s_{1/2}$ electron, the single active electron approximation (SAE) accurately describes their ionization dynamics, making them benchmark systems for theoretical investigations. Utilizing photoelectron spectroscopy, extensive theoretical~\cite{PhysRevA.104.L021103,PhysRevA.91.023424,toth2021probing,toth2021strong} and experimental~\cite{PhysRevLett.104.103003,schuricke2011strong,hart2016selective,zille2018chirp,wessels2018absolute,hockett2014complete} studies have been conducted on alkali metal atoms to investigate resonance-enhanced multiphoton ionization (REMPI) processes.

Previous research on the interaction between potassium atoms and intense laser fields has predominantly focused on laser intensities exceeding $\rm 10^{12}W/cm^{2}$. Due to the low ionization potential, the onset of over-the-barrier ionization (OBI) occurs at laser intensities several orders of magnitude lower than for hydrogen or noble gas atoms~\cite{PhysRevA.91.023424}. Specifically, the OBI threshold for a potassium atom is approximately $\rm 3.3\times10^{12}W/cm^{2}$\cite{PhysRevA.91.023424}. At laser intensities above this threshold, ionization of potassium atoms in near-infrared femtosecond laser fields exhibits multichannel characteristics, where several electronic states can resonate and contribute to the formation of free electron wave packets.

From a scattering perspective, photoionization can be regarded as a half-scattering process. By measuring photoelectrons, both the amplitude and phase of the emitted electron wave packets can be determined.
As the ionized electron traverses the effective Coulomb potential region generated by the nucleus and the residual electron, an additional phase is imparted to the final electron wave function. This phase is intricately connected to the underlying ionization dynamics and is directly correlated with the ionization time delay. Recent advances have enabled the extraction of this phase information, thereby enhancing the understanding of the temporal characteristics of ionization dynamics~\cite{doi:10.1126/science.1198450,doi:10.1126/science.1189401}. However, these investigations are complicated by additional phases introduced during measurements, which obscure the intrinsic nature of the observed phases~\cite{Azoury2019,doi:10.1126/science.aao7043}. 

In this paper, the photoelectron momentum distribution of potassium atoms from the photoionization by 800~nm femtosecond pulsed laser fields are studied. The AT-splitting caused by the near resonant one-photon transition of $4s$ ground state to $4p$ excited state is observed. Partial wave distributions of $p+f$ are analyzed in the momentum distribution. 
 Fitting to the experimental data allows us to determine the exact composition of the continuum wave function, and we discuss in details the dependence of the partial wave phase and amplitude with laser intensity.

\section{experimental methods}

We utilize an oven in our vacuum chamber to heat the metal potassium (to a temperature of 150~$^{\circ}$C, approximately) to produce an effusive atomic beam. The pressure of the vacuum chamber containing the atomic potassium source is about $10^{-5}$~Torr with the oven on. potassium vapor is jet out from a 0.2~mm hole at the front of the oven, and travels through a 2~mm diameter Skimmer and a 1~mm wide $\mu$-metal slit to interact with a linearly polarized femtosecond laser at the central wavelength of 800~nm. The repetition rate of the laser is 1~kHz, and the pulse duration is 50~$f$s. The laser polarization is kept linearly. The laser intensity is changed by rotating an 800~nm $\lambda$/2 wave plate combined with a polarizer. In our experiment, we employ a VMIs (velocity map imaging system) described in our previous papers~\cite{hu2019coherent,hu2020sub,Li_2021}, the photoelectrons are accelerated and focused to the MCP-PS (microchannel plates and phosphor screen) detector parallel to the laser polarization direction and the atomic beams. The photoelectrons are guided by the inhomogeneous electric field, which is generated by the repelling (with a typical applied voltage $V_{R}$ = $-2000$~V) and extraction ($V_{E}$ = $-1320$~V) electrodes to the detector. Photoelectron imaging is captured by a QSI camera and later processed on a personal computer.
The intensities of the laser used in our experiment are calibrated with the ponderomotive shift of the photoelectron of xenon atoms. Based on our previous work~\cite{Li_2021}, this calibration is further checked with the AT energy splitting of atomic potassium. 

\section{RESULT AND DISCUSSIONS}
\begin{figure}[t]
\begin{center}
\rotatebox{0}{\resizebox *{3.4in}{!} {\includegraphics {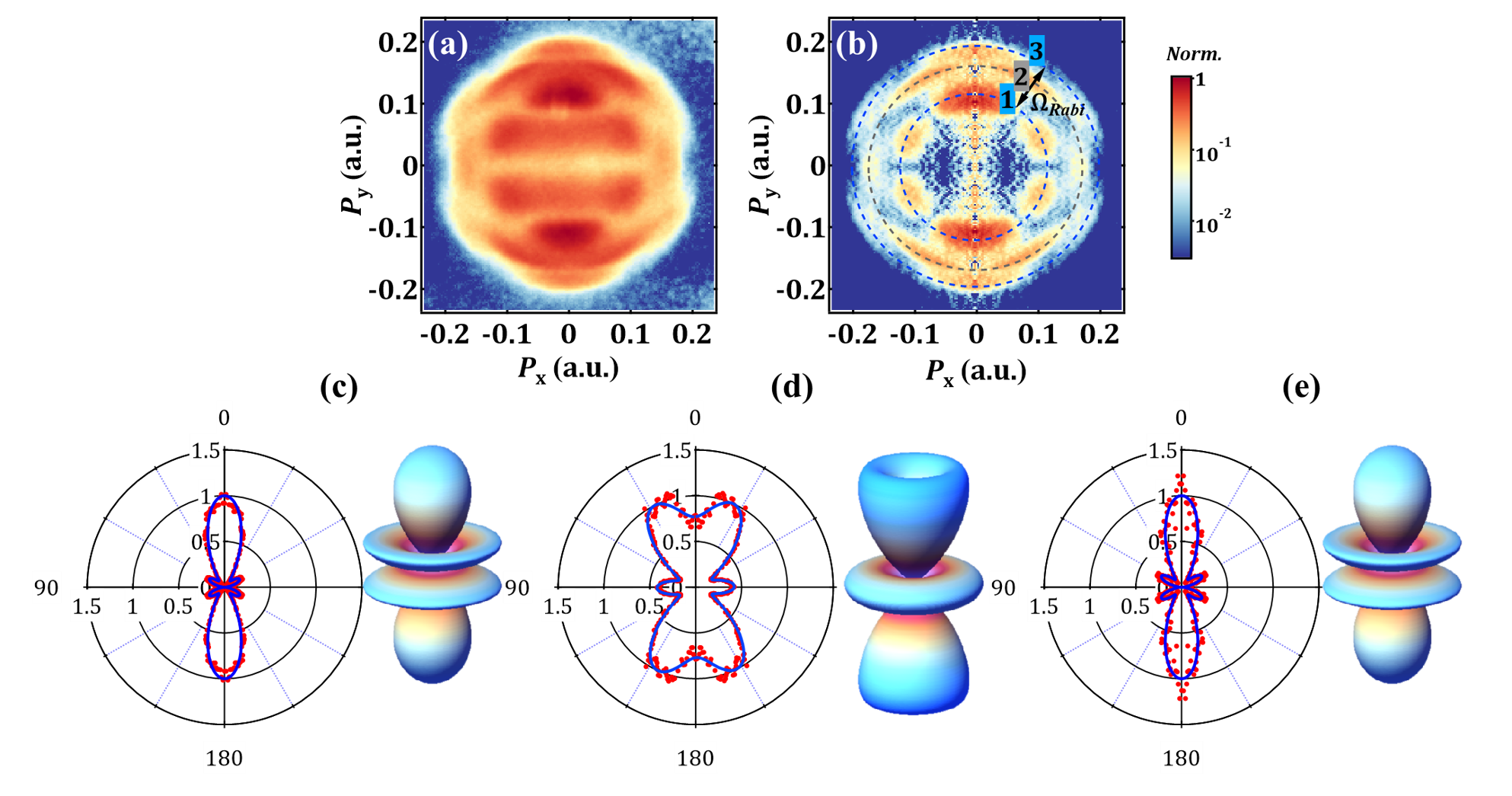}}}
\end{center}
\caption{Photoelectron momentum and angular distributions. (a) The original 2D photoelectron image acquired by the VMIs. (b) The reconstructed image using the Basex method~\cite{doi:10.1063/1.1807578,doi:10.1063/1.1482156}. There are three rings in the momentum distribution with distinct angular distributions shown in (c)-(e). The inner and outer most rings feature similar angular distributions. They present spherical-harmonics of the shape $Y_{lm}$, with $l$ = 3, and $m$ = 0. While the ring in the middle exhibits a $Y_{lm}$, with $l$ = 3, and $m$ = $\pm1$ structure, which is due to the spin-orbit interactions. (c)-(e) Are the angular distributions for each ring in the momentum distribution in (b). Red dots are experimental results and the blue lines are the fitting results. The corresponding spherical-harmonics are plotted along side for comparisons.}
\label{Fig. 1}  
\setlength{\abovedisplayskip}{1pt}
\setlength{\belowdisplayskip}{1pt}
\end{figure} 

 The measured photoelectron spectra are shown in Fig.~\ref{Fig. 1}. (a) represents the original image captured in the experiment and the image reconstructed by the inverse Abel transform (b) at the laser intensity of $\rm 1.2\times10^{11}~W/cm^{2}$, respectively. 
 The photoionization process investigated here are dominated by multiphoton ionization process as the calculated Keldysh parameter is much lager than 1. The potassium atom, with $I_{p}$ = 4.34~eV, can be ionized by absorbing three photons releasing the photoelectron with the kinetic energies of $E_{kin}$ = 3$\times$1.55~eV – $I_{p}$ = 0.31~eV. 
   
\begin{figure}[t]
\begin{center}
\rotatebox{0}{\resizebox *{3in}{!} {\includegraphics {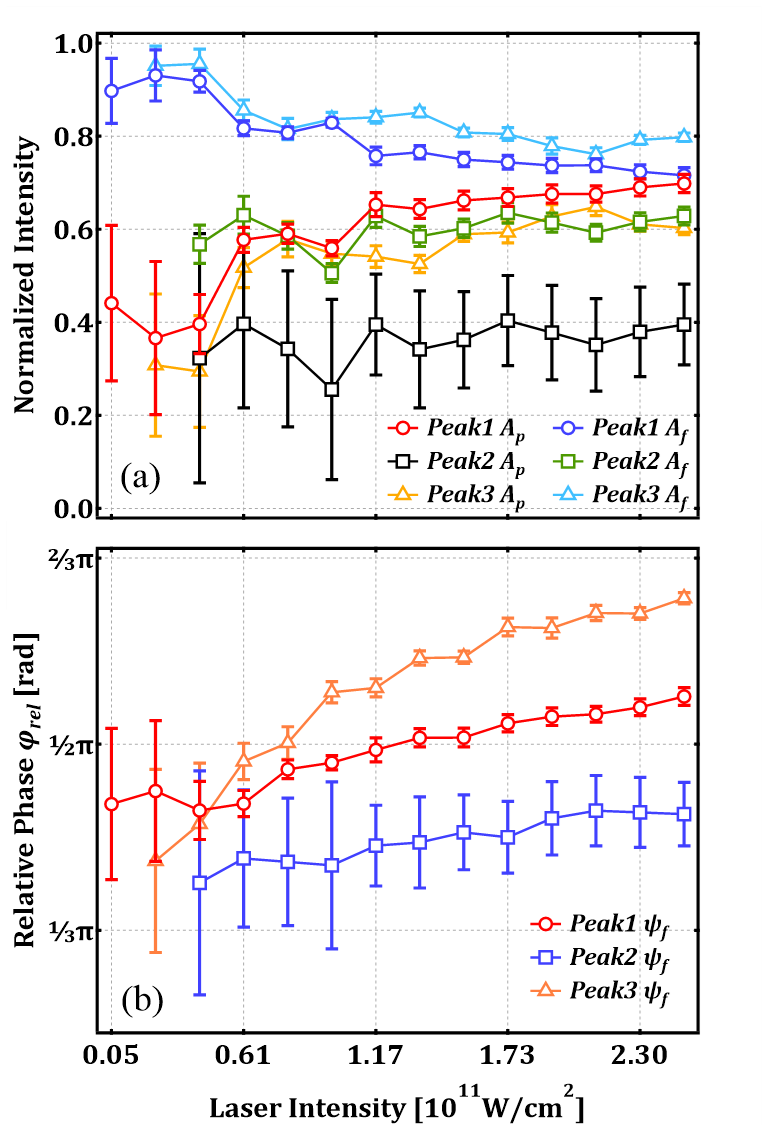}}}
\end{center}
\caption{Partial wave analysis. (a) represents the dependence of the partial wave amplitude of peak 1-peak 3 on laser intensity obtained by fitting Eq.~(\ref{angfit_1_3}), (b) represents the dependence of the partial wave phase obtained by fitting Eq.~(\ref{angfit_1_3}) on laser intensity.
%The wave amplitudes and phases of the three peaks tend to be flat near the laser intensity of $\rm 1.3\times10^{11}~W/cm^{2}$.
} 
\label{Fig. 2}  
\setlength{\abovedisplayskip}{1pt}
\setlength{\belowdisplayskip}{1pt}
\end{figure} 

The photoelectron kinetic energies $E_{kin}$ of the lowest three peaks exhibit distinct variations with increasing laser intensity. Specifically, the energy of peak 1 decreases in proportion to the square root of the laser intensity, while the energy of peak 2 remains constant. In contrast, the energy of peak 3 increases in proportion to the square root of the laser intensity.
The energy difference $\Delta E_{k}$ between peak 1 and peak 3 increase in proportion to the square root of the laser intensity, reaching 0.3~eV at an intensity of $\rm 1.9\times10^{11}~W/cm^{2}$. Such dependence of the energy difference between peak 1 and peak 3 with the laser intensities indicates that these two peaks are originated from the AT effect of the one-photon $4s$-$4p$ transition~\cite{autler1955stark,Li_2021}. 
The ionization channel of peak 2 is attributed to the spin-orbit interaction~\cite{bayer_time-resolved_2019}. 

\begin{figure}[t]
\begin{center}
\rotatebox{0}{\resizebox *{3.4in}{!} {\includegraphics {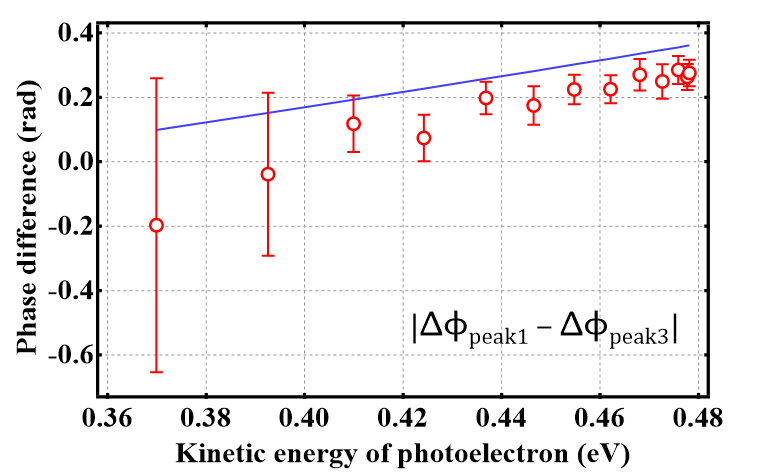}}}
\end{center}
\caption{The phase difference between the two AT-split peaks of the $4s-4p$ resonance is shown as a function of the kinetic energies of the photoelectrons. The red dots and line represent the experimentally measured values, while the blue line corresponds to the calculated phase difference, which includes contributions from the Coulomb phase and the quantum-defect phase.} 
 \label{Fig. 3}  
\setlength{\abovedisplayskip}{1pt}
\setlength{\belowdisplayskip}{1pt}
\end{figure}  
 
One-photon resonance results in the AT splitting can be described by:
\begin{equation}
\Delta E = \Omega=\frac{1}{\hbar}\left\langle a\left|er\right| b\right\rangle E_{0}
\label{two-photon}
\end{equation} 
where ${\hbar}$ is the reduced Planck constant, $|a\rangle$ and $|b\rangle$ are the two coupled states, $\left\langle a\left|er\right| b\right\rangle$ is the transition matrix element, and $E_{0}$ is the amplitude of the electric field of the laser~\cite{rickes2000efficient}. 

The bandwidth of our femtosecond laser pulse in this experiment is around 0.05~eV. In this condition, the $4s-4p$ transition can be driven near resonantly. The ionization channel of peak 1 and peak 3 is thus identified as a (1 + 2) REMPI from the ground state $4s$. As a result, the orbital quantum numbers (the partial waves) of peak 1 and peak 3 should be $l$ = 1, 3. The angular distribution of photoelectron shows a shape similar to the spherical harmonic function $Y_{30}$, which follows the Fano propensity rule~\cite{PhysRevA.32.617}. The angular part of the wave functions of peak 1 and peak 3 can be written as~\cite{villeneuve2017coherent,Boran_2021}:

  \begin{equation}
  \Psi(\phi,\theta,\psi)=A_{p} Y_{10}(\theta,\psi)+A_{f} e^{i \phi_{f}} Y_{30}(\theta,\psi)
  \label{angfit_1_3}
  \end{equation}where $A_{l}$ is the amplitudes of each partial wave contribution, $\phi_{l}$ is the corresponding partial wave phase. We set $\phi_{p} = 0$ to ensure only the relative phase are discussed. $Y_{lm}$ are spherical harmonics. %We use linearly polarized laser in our experiments, therefore the magnetic quantum number $m$ is a good quantum number and $m = 0$ for all considered partial waves.

Our measured PAD corresponds to the angular component of $|\psi|^{2}$. From Eq.~\ref{angfit_1_3}, it is evident that interference between different partial waves can be observed. This interference manifests in the angular distribution of photoelectrons, influencing features such as the nodal positions and the relative peak intensities.

Fig.~\ref{Fig. 2}(a) and (b) present the partial wave amplitudes and the relative phases of the angular distributions for peak 1 and peak 3 as functions of laser intensity. Here, $A_{p}$ and $A_{f}$ indicate the relative contributions of the channels with $\Delta l$ = 2. 
For both peaks, the partial wave amplitudes remain nearly constant over the examined laser intensity range. Notably, $A_{f}$ is larger than $A_{p}$, consistent with the Fano propensity rule~\cite{PhysRevA.32.617}. 

The dependence of the partial wave phase $\phi_{f}$ with the laser intensity is shown in Fig.~\ref{Fig. 2}(b). Unlike the intensity dependence of the partial wave amplitude, the phases of peak 1 and peak 3 changes dramatically with the laser intensity. Such phase shift can be related to the scattering process influenced by the residual charges during the escaping of the electron in ionization. In the case of non-resonant multiphoton ionization, the phase of the continuum photoelectron wave function of alkali metal atoms consists of two parts~\cite{wang2000determination}, the quantum-defect phase $\delta_{l}$, and the Coulomb phase\cite{burgess1960general,wang2000determination}, $\eta_{l}=\arg \{\Gamma[l+1-(i / \sqrt{2\varepsilon})]\}$. The quantum-defect phase of potassium atom is a slow varying function of photoelectron energy, and can be found from Ref.~\cite{Lorenzen_1983}. For the $p$ wave and $f$ wave of potassium this phase difference is given by $\delta_{f} - \delta_{p} = \left(1.874-0.744\varepsilon+1.75\varepsilon^{2}\right) \pi$, where $\epsilon$ is the kinetic energy of photoelectrons in atomic unit. 

% In our cases, the phase difference between $s$ and $d$ free electron partial waves of atomic potassium can be written in the following formula:
%     \begin{equation}
%    \begin{aligned}
%    \centering
%  \Delta\phi = &\phi_{s} - \phi_{d} = (\eta_{s} - \eta_{d}) + (\delta_{s} - \delta_{d}) 
% \\
%  = &\arctan \frac{1}{2 \sqrt{2\varepsilon}}+\arctan \frac{1}{\sqrt{2\varepsilon}} 
%  \\
%  &+\left(1.332-0.292\varepsilon-3.092\varepsilon^{2}\right) \pi
%  \end{aligned}
%  \label{Partial wave phase}
%  \end{equation} where $\epsilon$ is the kinetic energy of photoelectrons in atomic unit. 
%  Other effects which may significantly modify the phase shift of an ionized electron wave function include resonances, measurement procedure~\cite{PhysRevLett.46.1278,PhysRevA.27.861,PhysRevA.47.2881}. In our experiment we mainly focus on the lowest order photoelectron, and no continuum-continuum transition processes are involved as in the Rabbitt measurements~\cite{PhysRevA.92.063422}. 

In one-photon $4s-4p$ resonance, the electrons accumulate a large phase shift due to the resonance~\cite{PhysRevX.10.031070}. For peak 1 and peak 3, their photoelectrons originate from the same resonance state, the phase shift caused by the resonance effect is the same. 
Fig.~\ref{Fig. 3} shows the phase difference of the AT-splitting electrons together with the one obtained from Coulomb and quantum defect phase. The good agreement between the experimental results and the theoretical calculations demonstrates that the phase of the ionized electron can be attributed to the Coulomb phase plus the quantum defect phase. This phase is due to the scattering of the outgoing electron with the Coulomb potential from the ion core and the inner shell electrons. In fact, the $\Delta\phi$ of each peak reflects the scattering phase ~\cite{doi:10.1119/1.11721}, which can be related to Wigner time ~\cite{PhysRev.98.145}. Selecting an appropriate system to generate 1 + 1 REMPI can extract the time-delay between dressed states more accurately. Our results indicate that the angular distribution is an efficient tool to extract the phase information of ionized electrons , also served as an complementary tool comparing to other methods such as Rabbitt~\cite{doi:10.1126/science.1059413,doi:10.1126/science.1090277} or Attosecond streaking~\cite{PhysRevA.56.3870,hentschel_attosecond_2001,PhysRevLett.88.173903}. 

\section{conclusions}

In summary, we have studied the intensity-dependent energy spectrum and photoelectron angular distribution of potassium atoms under a near-infrared femtosecond laser pulse. We have analyzed and compared two different ionization channels from the one-photon induced AT-splitting. The splitting interval is proportional to the laser field strength. 
By carefully analyzing the angular distributions of the photoelectrons, we extract the relative phase of the continuum electrons, which we attribute to a combination of the Coulomb phase and the quantum defect phase. Our work provides a direct observation of one-photon resonance-induced AT-splitting, while also enabling coherent control of the atomic potassium wave packet. Additionally, we demonstrate that the angular distribution of the photoelectron serves as a powerful tool for obtaining phase information, offering a complementary method to study the fundamental timing properties in the ionization process.

\section{Acknowledgments}
   This work was supported by National Natural Science Foundation of China(Grant No. 12134005, and 12334011).  %D. Z. acknowledges the finical support of the starting grant from Jilin University.

\bibliography{references.bib}
\bibliographystyle{apsrev4-2}
\end{document}